\documentclass[runningheads]{llncs}
\usepackage{graphicx}

\usepackage{url}
\usepackage{amssymb}
\usepackage{amsmath}
\usepackage{multirow}
\usepackage{hyperref}
\usepackage{epstopdf}
\usepackage{caption}
\usepackage{subcaption}
\usepackage{algorithm}
\usepackage{algorithmic}
\usepackage{pgfplots}
\usepackage{CJKutf8}

\begin{document}
%
\title{Transfer Learning for Information Extraction with Limited Data}
%
%
\author{Minh-Tien Nguyen\inst{1,2} \and Viet-Anh Phan\inst{3} \and
Le Thai Linh\inst{1} \and
Nguyen Hong Son\inst{1} \and \\ Le Tien Dung\inst{1} \and Miku Hirano\inst{1} \and Hajime Hotta\inst{1}}
\authorrunning{Nguyen et al., published at PACLING 2019.}
%
\institute{CINNAMON LAB, \\10th floor, Geleximco building, 36 Hoang Cau, Dong Da district, Hanoi, Vietnam \\ \email{\{linhlt, levi, nathan, miku, hajime\}@cinnamon.is} \and
Hung Yen University of Technology and Education, Vietnam \\ \email{tiennm@utehy.edu.vn}\\
\and
Le Quy Don Technical University, Hanoi, Vietnam\\
\email{anhpv@mta.edu.vn}}
\maketitle              
\begin{abstract}
This paper presents a practical approach to fine-grained information extraction. Through plenty of authors’ experiences in practically applying information extraction to business process automation, there can be found a couple of fundamental technical challenges: (i) the availability of labeled data is usually limited and (ii) highly detailed classification is required. The main idea of our proposal is to leverage the concept of transfer learning, which is to reuse the pre-trained model of deep neural networks, with a combination of common statistical classifiers to determine the class of each extracted term. To do that, we first exploit BERT to deal with the limitation of training data in real scenarios, then stack BERT with Convolutional Neural Networks to learn hidden representation for classification. To validate our approach, {\bf we applied our model to an actual case of document processing using a public data of competitive bids for development projects in Japan.} We used 100 documents for training and testing and confirmed that the model enables to extract fine-grained named entities with a detailed level of information preciseness specialized in the targeted business process, such as a department name of application receivers.

\end{abstract}
\section{Introduction}\label{sec:intro}

The recent growth of AI technologies is dramatic, as many applications have already been deployed to real business cases \cite{mckinsey}. One of the most significant topics in the business scene is the conversion of unstructured text into structured one because it is stated as an entrance to the digital transformation \cite{william2007unstructured,digitaltransformation}. It has been studied by many proposed methods \cite{nlp}. In business services, the extraction of specific information matters such as organization names, personal names, addresses, or date plays an important role to facilitate document processing systems. Therefore, named entity recognition (NER) is one of the key technologies of natural language processing (NLP) which help AI software contributes to the commercial applications \cite{ner2}.

NER has been received attention due to its important role in many NLP applications \cite{lample2016neural,finkel2009nested,ju2018neural}. Conventional approaches use dictionary-based \cite{ner2}, which requires a huge amount of time to set up to keep it up-to-date. Recent approaches have been utilized machine learning to reduce the efforts of dictionary building and maintenance \cite{lample2016neural}. Majority of the researches have successfully done the simple tasks of NER, yet those have focused on relatively-easy tasks, e.g. extracting names of person or organizations; hence it is not always straightforward to apply the proposed algorithms to real cases due to two gaps. The first gap is about the amount of data. Document processing typically focuses on narrow, specific topics rather than general, wide ones, where only a small amount of annotated data is available. Let's take our case as an example. We only received 100 bidding documents for both training and testing. It challenges neural-network-based natural architectures, which usually require more than 10,000 annotated training examples. For instance, CoNLL 2003 provided 20,000 annotated words for NER \cite{devlin2018bert}. For actual cases, in a narrow domain, annotating the such number of training data is a time-consuming and non-trivial task. For the second gap, the identification of information types cannot be merely the categories \cite{fleischman2002fine,lee2006fine,shimaoka2016attentive}. The usual scope of NERs is to extract the categories of information such as organization names; however, in many cases, there can be two types of organizations, such as payee and payer in invoices. We believe that this level of detailed understanding is the key to the practical use of NER.

\begin{figure}[!h]
    \centering
  \includegraphics[width=9cm]{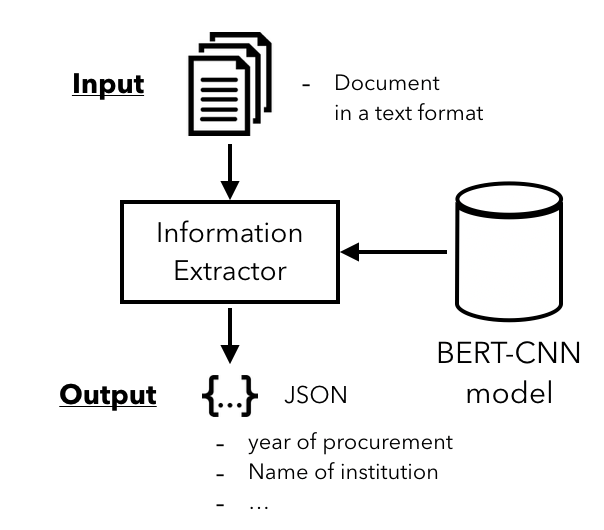}
  \caption{The overview of the system.}
  \label{fig:system}
\end{figure}

In this research, we aim to identify a particular type of named entities from documents with the limited amount of training data. The proposed model utilizes the idea of transfer learning to fine-grained NER tasks. As shown in Fig. \ref{fig:system}, the information extractor employs BERT \cite{devlin2018bert} for addressing the small number of training data (the first gap) and then stacks Convolutional Neural Network (CNN) with multilayer perceptron (MLP) to extract detail information, e.g. the year of procurement and the name of institution (the second gap). Outputs are stored in the JSON format for other tasks of document processing in our system. Our contributions are two-fold:
\begin{itemize}
    \item We propose a practical model of fine-grained NER which employs BERT \cite{devlin2018bert} as an element of transfer learning \cite{transferlearning}. By employing BERT and retraining the model with a small amount of data, our model achieves significant improvements compared to strong methods in extracting fine-grained NER. To the best our knowledge, we are the first study for the extraction of fine-grained NER for Japanese bidding documents.
    \item {\bf We present an applications to the actual case of competitive bids for development projects in Japan using a public data set and confirmed that the accuracy of the proposed model can be acceptably high enough for practical use. }
\end{itemize}

We applied our model to a real scenario of extracting information from bidding documents. Statistical analyses show that our model with small training data achieves improvements in term of F-score compared to strong baselines.

The rest of the paper is organized as follows: Section \ref{sec:related_work} presents the relevant research. Our proposed model is described in Section \ref{sec:the_approach}. Settings and evaluation metric are shown in Section \ref{sec:experiment}. We show the results, discussion, and error analysis in Sections \ref{sec:results} and \ref{sec:error-analysis}. Section \ref{sec:conclusion} concludes our investigations.

\section{Related Work}\label{sec:related_work}

Conventional methods utilize dictionaries \cite{watanabe2007graph} or take advantage of machine learning for named entity extraction. The dictionary-based method usually uses a pre-defined dictionary of entities to match tokens in documents. This method can achieve high accuracy, but it is time-consuming and labor-expensive to prepare the dictionary. In contrast, the machine learning method exploits features to train a classifier which can distinguish entities. This method has been shown efficiency for NER. Recently, the success of deep learning attracts researchers to apply this technique to information extraction. A recent study employs Long-Short Term Memory (LSTM) with a Conditional Random Field (CRF) to classify the contextual expressions \cite{lample2016neural}. More precisely, the authors used LSTM to learn the hidden representation of data and then stacked CRF for classification. This method showed promising results. In practice, several research projects focus on the nested named entities and have a great progress so far~\cite{finkel2009nested,ju2018neural}. 


For NER, high-level concepts such as people, places, organizations usually need to extract. However, for practical applications, categories must be in a more detailed level \cite{JJAndrew2018legal}. Here, fine-grained entity type classification was proposed, especially in the field of question answering, information retrieval \cite{lee2006fine,shimaoka2016attentive}, or the automatic generation of ontology \cite{fleischman2002fine}. However, the main challenge of fine-grained NER is the amount of training data required to train the classifier. To tackle this problem, transfer learning \cite{transferlearning} is an appropriate solution. It leverages pre-trained models trained by a large amount of out-domain data to build a new model with a small number of training data in a new domain. Transfer learning is efficient because we daily have faced with limited training data. Let's take our scenario as an example, we need to extract values for these tags in table 4 in long Japanese documents while the number of the training documents is only 78. Thus, transfer learning is one of the most efficient technique in such scenario \cite{shin2016deep,weiss2016survey}. 

One of the highlights of transfer learning is the recent open source named Bidirectional Encoder Representation from Transformers (BERT) \cite{devlin2018bert}. BERT is a form of transfer learning and has achieved state-of-the-art results on 11 NLP tasks, including the very competitive Stanford Question Answering Dataset (SQuAD). In this work, we develop a model based on BERT for our business task, which extracts information in long Japanese documents. 

\section{Proposed Approach} \label{sec:the_approach}
This section first introduces the problem and then describes our proposed model for information extraction from very long texts of Japanese bidding documents with limited training data.
\begin{figure}[!h]
    \centering
  \includegraphics[scale=0.45]{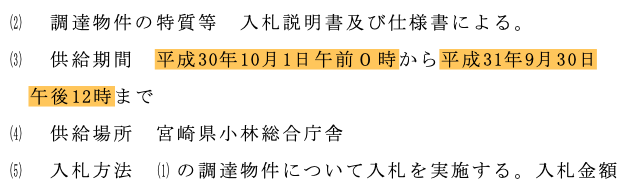}
    \caption{A part of bidding document. Clause 3 (two first lines) mean ''\textit{supply period from 04/01/2019 to 03/31/2020} (MM/dd/YYYY)".} 
\label{fig:bidding_doc_example}
\end{figure}

\subsection{Task Definition}
{\bf This research focuses on an actual case of document processing with a public actual data of competitive bids for development projects in Japan.} The task is to extract values for fields (tags) from bidding documents. A bidding document is a long document which has 3 sections 1) specifications, 2) invitation to bid, 3) instructions to bid.\vspace{-0.5cm}
\begin{table}[!h]
    \centering
  \caption{Information need to be extracted from bidding contract documents. Content type of name can be the name of an entity such as a company, department, or person.}\vspace{0.5cm}
  \includegraphics[width=9cm]{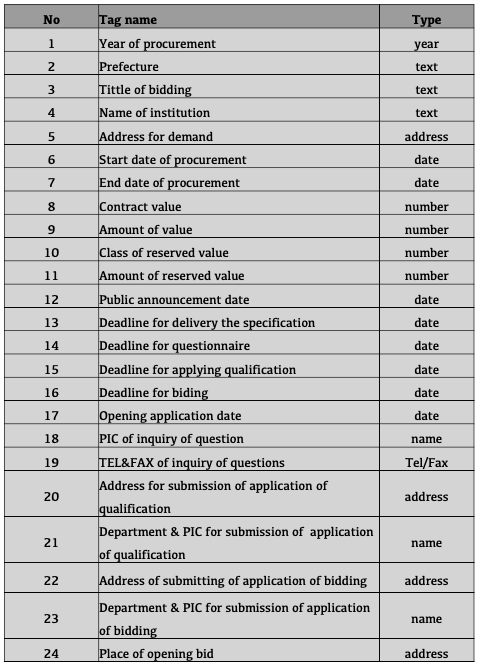}
\small
\label{table:tags}
\end{table}

Table \ref{table:tags} shows target information to be extracted. As observed, many tags may have the same value or data types, e.g. deadlines for the questionnaire and bidding. Hence, locating the value of a specific tag is challenging and needs to understand the context in the document.

\begin{figure}
  \includegraphics[width=8.2cm]{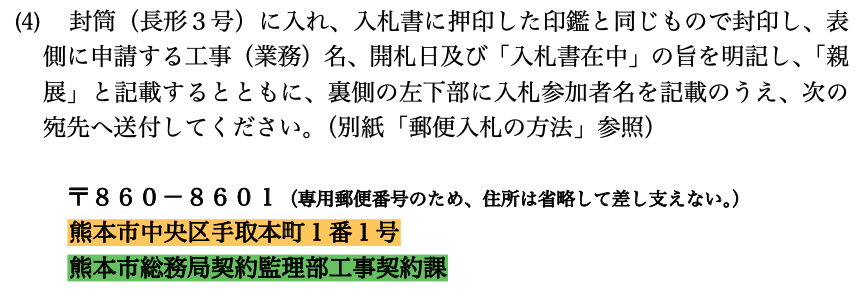}
    \caption{Example of a paragraph in a document.}
\label{fig:example}
\end{figure}

Fig. \ref{fig:bidding_doc_example} shows a part of a bidding document with three clauses 3, 4, 5 of the specification section. The yellow highlighted texts are the values for tag 6 (start date of procurement), and tag 7 (end date of procurement). Another difficulty is that a tag is not represented by specific terms. Therefore, to retrieve the values correctly, the learning model needs to understand the document structures and meaning of the text. For example, given a paragraph like Fig.~\ref{fig:example}, 
where the yellow line is the value of tags 20, 22, and the green line is the value of tags 18, 21, 23, 24, the model needs to decide which is the tag of the yellow and green lines correctly. To do that, we introduce a model which is the combination of BERT and CNN as follows.

\subsection{Proposed Model}\label{sec:model_architecture}
As mentioned, our model takes advantage of a pre-trained model (BERT), combined with CNN to learn the local context of each document for classification. Fig. \ref{fig:BERT_CNN} shows the overall architecture of the proposed model. The model has three main components: (i) the input vector representations of input tokens, (ii) BERT for learning hidden vectors for every token from the input tag and the document, and (iii) a convolution layer for capturing the local context and a softmax to predict the value location. The rest of this section will describe all parts of the model.

\subsubsection{Input representation}\label{sec:input_representation}
\begin{figure}[!h]
    \centering
  \includegraphics[width=12cm]{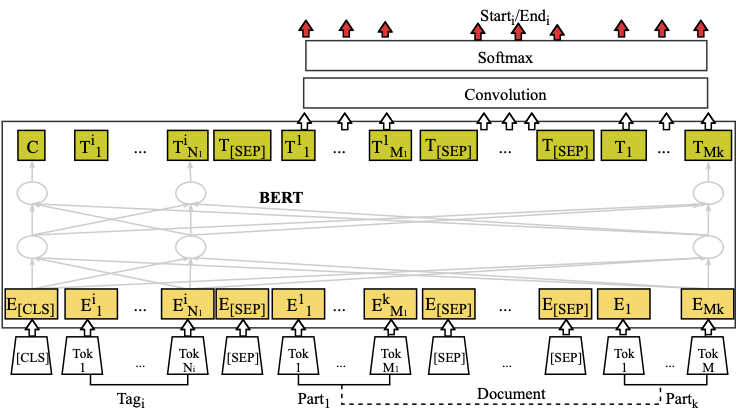}
  \caption{The BERT-CNN model for extracting the value of a tag.}
  \label{fig:BERT_CNN}
\end{figure}

Each input data includes the tag and the bidding document. The tag is treated as a single sequence, and the document is split into segments with a length of 384 tokens. Each token vector representation is determined based on the embeddings of token, sequence, and position. To differentiate among the input sequences, a special token (SEP) is inserted between them. For example, as showed in Fig. \ref{fig:BERT_CNN}, the tag $i$ has $N_i$ tokens, in which $token_1$ corresponds to embedding $E^i_1$. The document has $k$ sequences separated by token [SEP]. Most of the specification of embeddings follows the original paper \cite{devlin2018bert}.

\subsubsection{BERT}
BERT is a multi-layered bidirectional Transformer encoder, which allows our model to represent the context of a word by considering its neighbors \cite{devlin2018bert}. This is unlike unidirectional models that learn the contextual representation for each word using the words in one side (left or right). For example, considering the word ``bank" in two sentences ``I went to the bank to deposit some money" and "I went to the bank of the river", the representations of ``bank" are identical for a left-context unidirectional model, but they are distinguished with the use of BERT. This characteristic is compatible with our problem where many tags have homogeneous values. Let's take a look to Table \ref{table:tags}, given an address, it may be the address for demand (tag 5), submitting the application of qualification (tag 20), submitting for of bid (tag 22), or place of the opening bid (tag 24). Therefore, in our problem, the context aspect is critical to determine the tag that the address belongs to.

In our research, a pre-trained BERT was employed for two reasons. Firstly, BERT has shown state-of-the-art performance on many NLP tasks ranging over single/pair sentence classification, question answering, and sentence tagging. We, therefore, take advantage of BERT for our task of information extraction. Secondly, a pre-trained model is an appropriate solution to tackle our scenario of lacking training data. More precisely, we received only 100 documents for training and testing from our client. This is unpractical to train the whole network from scratch. To tackle this problem, we decided to use BERT as a type of transfer learning to fine-tune our model on bidding documents.

\subsubsection{Convolutional layer}\label{sec:cnn}
A convolutional layer was stacked on the top of BERT to learn local information of its input. Applying BERT is to take advantage of data in other domain, and produces hidden vectors from the tag and the bidding document. For this reason, the hidden vectors are for common texts that the network was pre-trained. To adapt to our domain, a convolutional layer is a essential component for capture statistical patterns. It retains the most important information of tags and extracted sequences in training data after convolution and pooling. This information is fed into the final layer for classification.

It is possible to use any neural network architecture to learn the hidden representation of tag and extracted text; however, we employed CNN because of two reasons. Firstly, it has been shown efficiency in capturing the hidden representation of data \cite{shin2016deep}. Secondly, our model needs to capture the local context of a tag and an extracted sequence in training data. By using CNN, our model can learn statistical patterns of data in a fast training process.

\subsubsection{Information extraction}\label{sec:information_extraction}
We formulate the information extraction problem as a question answering task. The value is pulled from the document by querying the tag. In the model, BERT learns the context of the document given the tag and produces hidden vectors for every token; the convolutional layer adjusts the vectors towards to our domain. Finally, a softmax layer is used to predict the location of the value. Each token is predicted to one of three outcomes including the start/end positions, and irrelevant to the tag. The extracted value is gathered based on the start to end positions with the highest probabilities.

\subsubsection{Training}\label{sec:training}
The training process of our model includes two stages: (i) pre-training and (ii) fine-tuning. For the first stage, the pre-trained weights of BERT were reused, while the weights of the rest layers were randomly initialized. The BERT was trained with a large text corpus of Japanese collected from Wikipedia. The training task is to predict whether a sentence is the next or just a random of other sentence \cite{devlin2018bert}. The setting is the same as the original paper \cite{devlin2018bert}.

The whole network was fine-tuned in 20 epochs with our training data by using cross-entropy. We used multilingual BERT-based model trained for 102 languages on a huge amount of texts from Wikipedia. The model has 12 layers, a hidden layer of 768 neurons, 12 heads and 110M parameters. Convolution uses 768 filters with the window size = 3.

\section{Settings and Evaluation Metric}\label{sec:experiment}

\subsection{Dataset}
{\bf The dataset contains 100 bidding documents}, in which 78 documents were used for training and 22 documents for testing. From Tables \ref{table:tags} and \ref{table:datastatistics}, we can observe that the extraction extracts many similar types of short values in a very long document, which has an average of 616 sentences.

\begin{table}[!h]
\centering
\caption{Statistics of the dataset.}
\begin{tabular}{lcc}
\hline
Statistics & mean & Std. \\ \hline
\#training samples & 78 & - \\
\#testing samples & 22 & - \\
\#characters/sample & 22,537 & 17,191 \\
\#sentences/sample & 616 & 368 \\ \hline
\end{tabular}
\label{table:datastatistics}
\end{table}

\subsection{Baselines}
To verify the efficiency of our approach, we compare our model to five baselines as follows.
\begin{itemize}
    \item \textbf{BERT}: is the basic model which obtains state-of-the-art performance on many natural language processing tasks, including QA \cite{devlin2018bert}. We directly applied BERT for QA on our testing data, without any additional training.
    
    \item \textbf{BERT+CRF}: stacks CRF on BERT for prediction. This is because CRF is a conventional method for information extraction and NER. This method was trained on training data and then applied to test data.
    
    \item \textbf{BERT+LSTM+CRF}: This model uses LSTM to capture the hidden representation of sequences and employs CRF for classification. This is a variation of LSTM-CRF for NER \cite{lample2016neural}. We also tried with BiLSTM but its results are not good to report.
    
    \item \textbf{$n$-grams+MLP+regex}: was trained with $n$-gram features ($n$ in $[1, 4]$). The MLP was used to predict the paragraph containing the values of tags. The regular expression was finally applied to extract the values. 
    
    \item \textbf{Glove+CNN+BiLSTM+CRF}: We used a deep neural network including layers of convolution, bidirectional Long Short Term Memory (BiLSTM), and conditional random fields (CRFs) to automatically extract tags of input texts. For token embedding, we use Glove \cite{pennington2014glove}.

\end{itemize}

\subsection{Evaluation metric}
We used F-score (F-1) to evaluate the performance of our model as well as the baselines. We matched extracted outputs to ground-truth data to compute precision, recall, and F-score. The final F-score was computed on all tags.

\section{Results and Discussion}\label{sec:results}
Table \ref{table:f1_scores} reports the comparison according to F1-score. On average, our method outperforms others notably. This is because our model exploits the efficiency of BERT trained on a large amount of data; therefore, it can potentially capture the hidden representation of data. By stacking convolution and retraining the model with 78 training documents, our model has the ability to adapt to a new domain. As a result, the model can correctly extract information on bidding documents and can improve the performance over BERT-QA of 4.55\%. The improvements come from two possible reasons: (i) we employ BERT to tackle the limited number of training data and (ii) we take advantage of CNN for fine-tuning to capture local context. This confirms our assumption in Section \ref{sec:intro} that we can utilize transfer learning for information extraction in a narrow domain with limited data.

\begin{table}[ht!]
\centering
\caption{Comparison of methods according F1-score}
\begin{tabular}{lc}
\hline
Method & Avg. of F1 \\ \hline
BERT (QA)  & 0.8607 \\
BERT+CRF & 0.1773 \\
BERT+LSTM+CRF & 0.3817 \\
BERT+CNN+MLP (Our model)  &\textbf{ 0.9062} \\ \hline
$n$-grams+MLP+Regex & 0.8523 \\
CNN+BiLSTM+CRF+MLP & 0.6766\\ \hline
\end{tabular}
\label{table:f1_scores}
\end{table}

The BERT-QA model is the second best. It is understandable that this model also uses BERT as a pre-trained model to extract hidden representation. However, it needs to be adapted to a new domain. It shows the efficiency of retraining the model on a new domain. Even with a small amount of data (78 training documents), our model can improve the F-score. Interestingly, BERT+CRF and BERT+LSTM+CRF do not show improvements. For BERT-CRF, it may lack the patterns learned from training data because it directly uses outputs from BERT for CRF to do classification. For BERT-LSTM-CRF, the possible reason is bidding documents are very long (Table \ref{table:datastatistics}); therefore, long-term dependencies may affect these models. The model using MLP with $n$-grams features achieves competitive results. This is because this model uses two steps to extraction candidates. The first step uses MLP to detect whether a sentence contains extracted information or not. The second step uses regular expressions to extract candidates. However, in some cases, the expressions cannot cover the patterns of text (please refer to Section \ref{sec:error-analysis}). An interesting point is that the model CNN+BiLSTM+MLP outputs low scores, in which its performance is lower than $n$-grams+MLP of 17.57\%. As mentioned above, training deep neural networks from scratch with limited data is difficult to convergence because we need to optimize a large number of parameters. The feature-based method requires less training data but it easily suffers from the definition of regular expressions.

\begin{figure}[ht!]
    \centering
  \includegraphics[width=11.5cm]{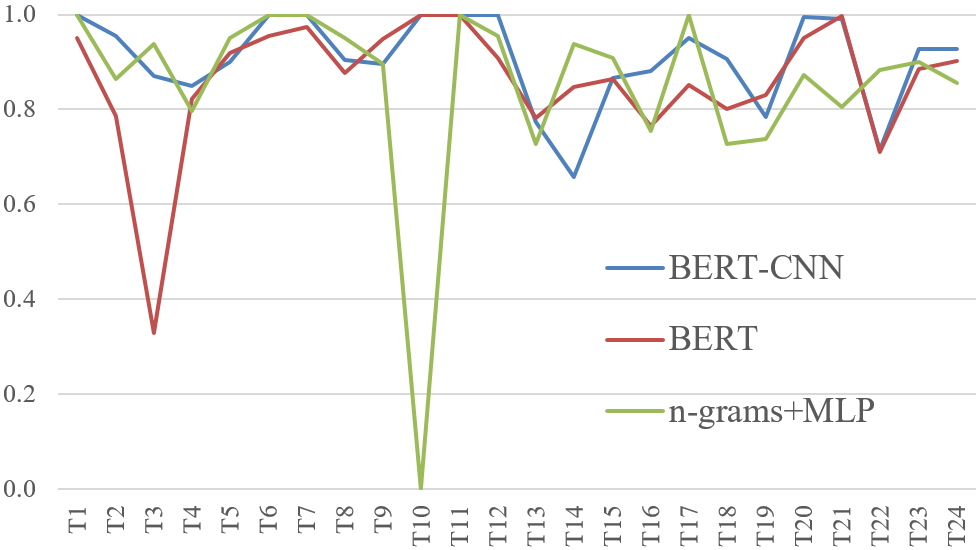}
  \caption{F1-scores per tag.}
  \label{fig:f1_scores}
\end{figure}

We observed F-score on each tag to analyze how each model works. To do that, we matched extracted text to reference and compute scores, which are plotted in Fig. \ref{fig:f1_scores}. We did not plot the results of BERT+CRF, BERT+LSTM+CRF, and CNN due to its low scores.
As can be seen, our model works stably on all tags, and usually achieves the best F1 scores in comparison with BERT-QA and $n$-grams-MLP. This again supports results in Table \ref{table:f1_scores} where our model is the best. {\bf It should be noted that the feature-based method obtains 0\% of F1 for tag 10 (classification of the reserved values related to the case). The reason is that $n$-grams features and the regular expressions may not capture well the patterns of these values (tag 10)}. This shows that the $n$-grams+MLP+regex method can be affected by the definition of regular expressions. As the result, it could not find any values for tag 10 on the test set. By contrast, our model trained on BERT can exploit rich hidden representation from large pre-trained data and with 78 training documents of bids, the model can correctly extract tag 10.


\section{Error Analysis}\label{sec:error-analysis}
In this section, we investigate how our model works on real data of bids. We observed the output of our model and the baselines. Table \ref{table:error_analysis} shows some predicted values for three tags 21, 22, 23 (\ref{table:tags}) in documents of two contracts. The first part shows a case that our model do not work and the second part presents a case that our model extracts correct sequences.
\begin{CJK}{UTF8}{min}
\begin{table}[]
\caption{Some examples of model outputs. The correct answers and correct predictions are highlighted in blue. We did not show outputs of BERT, BERT+CRF and BERT+LSTM+CRF due to space limitation.}
\label{table:error_analysis}
\begin{tabular}{p{1cm} p{11cm}}
\hline

\textbf{File} & \textbf{Sample A} \\ \hline

Tag & \multicolumn{1}{c}{\textbf{Correct answer}} \\ \hline
21 & {\color[HTML]{3531FF} 熊谷市東区役所東町支所1階総務民生課} (Kumagaya City Higashi Ward Office Higashimachi Branch 1F General Affairs and Welfare Division). \\ 
22 & {\color[HTML]{3531FF} 〒23－583東町郵便局留} (Higashi-machi Post Office Station, postal code: 23－583). \\ 
23 & {\color[HTML]{3531FF} 熊谷市役所東区役所東町支社宛} (Kumagaya City Hall Higashi Ward Office Higashi-machi Branch Office). \\ \hline

 & \multicolumn{1}{c}{\textbf{n-gram MLP}} \\ \hline
21 & {\color[HTML]{3531FF} 熊谷市東区役所東町支所1階総務民生課} \\ 
22 & {\color[HTML]{3531FF} 〒23－583東町郵便局留} \\ 
23 & 熊谷市役所東区役所東町支社宛 \\ \hline 

 & \multicolumn{1}{c}{\textbf{CNN+BiLSTM}} \\ \hline
21 & 熊谷市東区東町東45番地熊谷市東区役所東町支所宛 \\ 
22 & 〒23-585東町郵便局留熊谷市役所東区役所東町支所宛 \\ 
23 & 〒23-585東町郵便局留熊谷市役所東区役所東町支所宛 \\ \hline

 & \multicolumn{1}{c}{\textbf{Our model}} \\ \hline
21 & 東町支所総務民生課 \\ 
22 & 熊谷市東区東町東45番地 \\ 
23 & 東町支所総務民生課 \\ \hline \hline

\textbf{File} & \textbf{Sample B} \\ \hline

Tag & \multicolumn{1}{c}{\textbf{Correct answer}} \\ \hline
21 & {\color[HTML]{3531FF} 新潟県会計管理局物品管理調達課物品調達担当}  (Niigata Prefecture Accounting Administration Department Goods Management Procurement Section Goods Procurement). \\ 
22 & {\color[HTML]{3531FF} 〒11-878新潟市小路2丁目1番地10} (postal code 11-878 Niigata, Niigata, Koji, 2 Chome−1−10). \\ 
23 & {\color[HTML]{3531FF} 県立新潟病院総務課整備担当} (Prefectural Niigata Hospital General Affairs Division Maintenance Section). \\ \hline

 & \multicolumn{1}{c}{\textbf{n-gram MLP}} \\ \hline
21 & (232)6118 \\ 
22 & 目1番地10郵便番号11-878電話番号10 \\ 
23 & (19)618 \\ \hline 

 & \multicolumn{1}{c}{\textbf{CNN+BiLSTM}} \\ \hline
21 & 目1番地10郵便番号11－ \\ 
22 & 新潟市小路2丁 \\ 
23 & 新潟市小路2丁 \\ \hline

 & \multicolumn{1}{c}{\textbf{Our model}} \\ \hline
21 & {\color[HTML]{3531FF} 新潟県会計管理局物品管理調達課物品調達担当} \\ 
22 & {\color[HTML]{3531FF} 〒11-878新潟市小路2丁目1番地10} \\ 
23 & {\color[HTML]{3531FF} 新潟県会計管理局物品管理調達課物品調達担当} \\ \hline

\end{tabular}
\end{table}
\end{CJK}

After observing, we found that for the feature-based method, it may suffer from the out of vocabulary problem because regular expressions were created by on keywords defined by humans. Although the classifier can predict correctly the sentence containing the value, the regular expression fails to locate the information. Taking the contract "Sample B" as an example, since the address of the company is not in the training data, the regular expression only captures the number part. Unlikely, the deep models can predict both the text and number parts in this case. However, the model trained from scratch usually produces uncompleted text, especially with the long-text values. This comes from the lack of training data that is one of the essential keys to allow the convergence of a deep model. In contrast, the pre-trained model can predict the whole content.

We also found that the feature-based methods can predict short values like date times, and numbers efficiently. In this case, the classifier finds the sentence using features and the regular expression can locate the information easily. For these types of fields, the performance of the feature-based and the deep models is similar. This is shown in Fig. \ref{fig:f1_scores}.

\section{Conclusion}\label{sec:conclusion}
We presented a transfer learning method for information extraction with limited data. In the experiment, we used 100 documents for the training and validation, and confirmed that our method achieves high accuracy of fine-grained classification with the limited number of annotated data. We believe that the result implies a positive aspect of the concept of transfer learning for real information extraction scenarios.

A possible direction of our research is to investigate more final layers for classification. We also encourage to add heuristic features generated from humans and change the structure of our model by replacing CNN by more deeper structures, e.g. CNN-BiLSTM.

\section*{Acknowledgement}
We would like to thank to NLP and publication board members of Cinnamon Lab for useful discussions and insightful comments on earlier drafts. We also thank to anonymous reviewers for their comments for improving our paper. 


%
%
%
\bibliographystyle{splncs04}
\bibliography{pacling}
%




\end{document}